\documentclass[12pt]{article}
\usepackage{amsmath}
\usepackage{graphicx}
\usepackage{subfigure}
\usepackage{hyperref}

\begin{document}

\title{Interacting tachyon Fermi gas}
\author{Ernst Trojan \and \textit{Moscow Institute of Physics and Technology} \and 
\textit{PO Box 3, Moscow, 125080, Russia}}
\maketitle

\begin{abstract}
We consider a system of many fermionic tachyons coupled to a scalar,
pseudoscalar, vector and pseudovector fields. The scalar and pseudoscalar
fields are responsible for the effective mass, while the pseudovector field
is similar to ordinary electromagnetic field. The action of vector field $%
\omega _\mu $ results in tachyonic dispersion relation $\varepsilon _p=\sqrt{%
p^2+g^2\omega _0^2-hpg\omega _0-g\vec \sigma \cdot \nabla \omega _0-m^2}%
-g\vec \sigma \cdot \vec \omega $ that depends on helicity $h$ and spin $%
\vec \sigma $. We apply the mean field approximation and find that there
appears a vector condensate with finite average $\left\langle \omega
_0\right\rangle $ depending on the tachyon density. The pressure and energy
density of a many-tachyon system include the mean-field energy $\left\langle
\varepsilon _p\right\rangle =\sqrt{p^2+hpng^2/M^2+n^2g^4/M^4-m^2}$ which is
real when the particle number density exceeds definite threshold which is $%
n>mM^2/g^2$ for right-handed and $n>\frac 2{\sqrt{3}}mM^2/g^2$ for
left-handed tachyons, while all tachyons are subluminal at high density.
There is visible difference in the properties of right-handed and
left-handed tachyons. Interaction via the vector field $\omega _0$ may lead
to stabilization of tachyon matter if its density is large enough.
\end{abstract}


\section{Introduction}

Tachyon is a substance whose group velocity is larger than the speed of
light in vacuum $c=1$. Tachyons are commonly known in the field theory \cite
{C1} and nonlinear optics \cite{O1}. A free fermionic tachyon is described
by the Lagrangian \cite{D1,D2} 
\begin{equation}
L=\bar \psi \left( i\gamma _5\gamma ^\mu \partial _\mu -m\right) \psi
\label{la0}
\end{equation}
corresponding to the tachyonic Dirac equation $(i\gamma ^\mu \partial _\mu
-\gamma _5m)\psi =0$ that has plane-wave solution in the form 
\begin{equation}
\psi =\left( 
\begin{array}{c}
\phi \\ 
\chi
\end{array}
\right) \exp \left( i\vec p\cdot \vec r-i\varepsilon _{p\,}t\right)
\label{wa}
\end{equation}
with a single-particle tachyon energy 
\begin{equation}
\varepsilon _p=\sqrt{p^2-m^2}  \label{tah}
\end{equation}

Much more complicated theory is developed for tachyons in external fields 
\cite{Da98,K2011}. If a charged Dirac tachyon interacts with electromagnetic
field $A_\mu $, the Lagrangian is presented in the form \cite{D3} 
\[
L=\bar \psi \left( i\gamma _5\gamma ^\mu D_\mu -m\right) \psi 
\]
where $D_\mu =\partial _\mu +iA_\mu $. Dirac tachyons with nonlinear
self-interaction were also considered \cite{O2011}. However a system of many
interacting tachyons has not been discussed yet.

In the present paper we consider the tachyonic Lagrangian 
\begin{eqnarray}
&L=\bar \psi \left( i\gamma _5\gamma ^\mu \partial _\mu -m+s+i\gamma _5\Pi
-g\gamma ^\mu \omega _\mu -\gamma _5\gamma ^\mu A_\mu \right) \psi +\frac
12\left( \partial ^\nu s\partial _\nu s-M_s^2s^2\right) +&  \nonumber \\
&+\frac 12\left( \partial ^\nu \Pi \partial _\nu \Pi -M_\pi ^2\Pi ^2\right)
+\frac 12\left( G^{\mu \nu }G_{\mu \nu }+M^2\omega _\mu ^2\right) +\frac
12\left( F^{\mu \nu }F_{\mu \nu }+M_A^2\,A_\mu ^2\right) &  \label{la}
\end{eqnarray}
that includes interaction with scalar field $s$, pseudoscalar field $\Pi $,
massive vector field $\omega _\mu $ (relevant Maxwell tensor is $G_{\mu \nu
}=\partial _\mu \omega _\nu -\partial _\nu \omega _\mu $) and massive
electromagnetic field $A_\mu $ ($F_{\mu \nu }=\partial _\mu A_\nu -\partial
_\nu A_\mu $). Each interaction vertex in (\ref{la}) is taken in the form of
Yukawa minimal coupling while the coupling constants are included in the
fields for simplicity. Our interest concerns the single-particle energy
spectrum and a the thermodynamical functions of many tachyon system in the
frames of the mean-field approach. The interaction may result in a field
condensate with non-zero vacuum expectation value responsible for
qualitatively new properties in contrast to a free tachyon Fermi gas.

\section{Tachyonic energy spectrum}

Starting with the Lagrangian (\ref{la}), we write the equation of motion for
the Dirac tachyon 
\begin{equation}
\left( i\gamma ^\mu \partial _\mu -\gamma _5m+\gamma _5s+i\Pi -\gamma
_5\gamma ^\mu \omega _\mu -\gamma ^\mu A_\mu \right) \psi =0  \label{dir}
\end{equation}
and Klein-Gordon equations 
\begin{equation}
\partial ^\mu \partial _\mu s+M_s^2s=\bar \psi \psi  \label{ss}
\end{equation}
\begin{equation}
\partial ^\mu \partial _\mu \Pi +M_\pi ^2\Pi =\bar \psi \gamma _5\psi
\label{pi}
\end{equation}
\begin{equation}
\partial ^\mu G_{\mu \nu }+M^2\omega _\mu =\bar \psi \gamma _\mu \psi
\label{oo}
\end{equation}
\begin{equation}
\partial ^\mu F_{\mu \nu }+M_A^2A_\mu =\bar \psi \gamma _\mu \gamma _5\psi
\label{aa}
\end{equation}
for the fields $s$, $\Pi $, $\omega _\mu $ and $A_\mu $, where the tachyons
are acting as sources.

Looking for solution of equation (\ref{dir}) in the form of plane wave (\ref
{wa}), we come to a linear system for bispinors 
\begin{equation}
\begin{array}{c}
\left[ \vec \sigma \cdot \left( \vec p-\vec A\right) +m-s-\omega _0\right]
\chi =\left( \varepsilon _p+\vec \sigma \cdot \vec \omega +A_0+i\Pi \right)
\phi \\ 
\left[ \vec \sigma \cdot \left( \vec p-\vec A\right) -m+s-\omega _0\right]
\phi =\left( \varepsilon _p+\vec \sigma \cdot \vec \omega +A_0-i\Pi \right)
\chi
\end{array}
\label{bi}
\end{equation}
It has solution if the energy satisfies dispersion relation 
\begin{equation}
(\varepsilon _p+\vec \sigma \cdot \vec \omega +A_0)^2=\left[ \vec \sigma
\cdot (\vec p-\vec A)-\omega _0\right] ^2-m_{*}^2  \label{ea}
\end{equation}
where 
\begin{equation}
m_{*}^2=\left( m-s\right) ^2+\Pi ^2  \label{em}
\end{equation}
is the effective mass. Indeed, switching off all interactions, we obtain
from (\ref{ea}) the energy spectrum of a free tachyon (\ref{tah}).

Taking into account the properties of the Dirac matrices $\gamma _\mu $ and
operator $\vec p=-i\nabla $, we find that the energy spectrum (\ref{ea}) at $%
\omega _\mu =0$ is reduced to 
\begin{equation}
\varepsilon _p=\sqrt{\left[ \vec \sigma \cdot \left( \vec p-e\vec A\right)
\right] ^2-m_{*}^2}-eA_0=\sqrt{\left( \vec p-e\vec A\right) ^2-e\vec \sigma
\cdot \mathrm{rot}\vec A-m_{*}^2}-eA_0  \label{el}
\end{equation}
where the coupling constant $e$ is written explicitly so that expression (%
\ref{el}) is looking like well known energy spectrum in electrodynamics and
nuclear physics \cite{KaM} where potential $A_\mu $\ is acting as an
ordinary electromagnetic or vector-meson filed, while the scalar field $s$
and pseudoscalar field $\Pi $ are responsible for the effective mass \ref{em}%
.

When the vector field $\omega _\mu \neq 0$ is switched on, an absolutely
different tachyonic energy spectrum is obtained: 
\begin{equation}
\varepsilon _p=\sqrt{p^2+g^2\omega _0^2-g\omega _0\vec \sigma \cdot \vec
p-g\vec \sigma \cdot \nabla \omega _0-m^2}-g\vec \sigma \cdot \vec \omega
\label{ew}
\end{equation}
Here we omit for simplicity all other fields ($s$, $\Pi $, $A_\mu $) and
write the coupling constant $g$ explicitly. There is not only dependence on
the spin (note that $\nabla \omega _0$ is acting as a regular magnetic field
applied to an ordinary non-tachyon particle) there is also explicit
dependence on the helicty of the particle.

For uniform electro-static potential $\omega _0=\mathrm{const}$ (and $\vec
\omega =0$) formula (\ref{ew}) is reduced to 
\begin{equation}
\varepsilon _p=\sqrt{p^2+g^2\omega _0^2-hg\omega _0p-m^2}  \label{ew2}
\end{equation}
where $h=1$ for right-handed and $h=-1$ left-handed tachyons, respectively.
The energy (\ref{ew2}) of right-handed tachyons is imaginary at any $p$ if 
\begin{equation}
-m<g\omega _0<\frac 2{\sqrt{3}}m  \label{rr}
\end{equation}
The energy (\ref{ew2}) of left-handed tachyons is imaginary at any $p$ if 
\begin{equation}
-\frac 2{\sqrt{3}}m<g\omega _0<m  \label{ll}
\end{equation}

The group velocity 
\begin{equation}
v=\frac{d\varepsilon _p}{dp}=\frac{p-\frac 12hg\omega _0}{\sqrt{%
p^2+g^2\omega _0^2-hg\omega _0p-m^2}}  \label{vel}
\end{equation}
is subluminal at any $p$ if 
\begin{equation}
\left| g\omega _0\right| >\frac 2{\sqrt{3}}m  \label{gr}
\end{equation}

\section{Mean field approach to many-tachyon system}

Now consider a many-tachyon system with interaction described by the
Lagrangian (\ref{la}) and enclosed in volume $V=\int d^3r$. A charge $Q$,
associated with the relevant Noether current, is incorporated in the
partition function \cite{KaM} 
\begin{equation}
Z=\int \left[ d\bar \psi \right] \left[ d\psi \right] \exp \left\{
\int\limits_0^{1/T}d\tau \left( \int Ld^3r+\mu Q\right) \right\}  \label{z}
\end{equation}
where $\tau =it$, and $\mu $ is the tachyonic chemical potential. The
tachyonic equation of motion (\ref{dir}) results in the continuity equation $%
\partial _{\mu \,}j_5^\mu =0$ for the axial current $j_5^\mu =\bar \psi
\gamma ^\mu \gamma _5\psi $ that corresponds to the conserved axial charge $%
Q_5=\int j_5^0d^3r$, while the vector current $j^\mu =\bar \psi \gamma ^\mu
\psi $ is not conserved because \cite{T89} $\partial _{\mu \,}j^\mu
=-2im\bar \psi \gamma _5\psi $. Hence, the axial charge is incorporated in
the partition function (\ref{z}) rather than the number of particles (which
is not conserved). Thus, the tachyonic pressure 
\begin{equation}
P=T\ln Z  \label{ppp}
\end{equation}
the energy density 
\begin{equation}
E=\mu n_5+TS-P  \label{eee}
\end{equation}
and the entropy density 
\begin{equation}
S=\left( \frac{\partial P}{\partial T}\right) _{V,\mu }  \label{eee}
\end{equation}
include the axial charge density 
\begin{equation}
n_5=T\frac{\partial \left( \ln Z\right) _{V,T}}{\partial \mu }  \label{n}
\end{equation}
rather than ordinary particle number density.

Let us apply the mean field approximation \cite{KaM}, neglecting quantum
field fluctuations. The tachyons are taken to move independently in the
presence of constant mean fields $\left\langle s\right\rangle $, $%
\left\langle \Pi \right\rangle $, $\left\langle \omega _\mu \right\rangle $
and $\left\langle A_\mu \right\rangle $ which themselves are generated
self-consistently according to equations (\ref{ss})-(\ref{aa}), namely: 
\begin{equation}
M_s^2\left\langle s\right\rangle =\left\langle \bar \psi \psi \right\rangle
\label{ss1}
\end{equation}
\begin{equation}
M_\pi ^2\left\langle \Pi \right\rangle =\left\langle \bar \psi \gamma _5\psi
\right\rangle  \label{pi1}
\end{equation}

\begin{equation}
M^2\left\langle \omega _\mu \right\rangle =\left\langle \bar \psi \gamma
_\mu \psi \right\rangle   \label{oo1}
\end{equation}
\begin{equation}
M_A^2\left\langle A_\mu \right\rangle =\left\langle \bar \psi \gamma _\mu
\gamma _5\psi \right\rangle   \label{aa1}
\end{equation}
The argument of partition function (\ref{z}) is 
\[
\bar L+\tilde L+\mu Q_5
\]
where $\bar L$ is the Lagrangian of tachyons and 
\[
\tilde L=-\frac 12M_s^2\left\langle s\right\rangle ^2-\frac 12M_\pi
^2\left\langle \Pi \right\rangle ^2+\frac 12M^2\left\langle \omega _\mu
\right\rangle ^2+\frac 12M^2\left\langle A_\mu \right\rangle ^2
\]
is the Lagrangian of the mean fields. It allows to present the partition
function (\ref{z}) as a product $Z\cong \left\langle Z\right\rangle =\bar
Z\tilde Z$ that includes contribution of the tachyons $\bar Z$ and
contribution of the mean fields $\tilde Z$. Thus, we determine the pressure 
\begin{equation}
\left\langle P\right\rangle =\bar P-\frac 12M_s^2\left\langle s\right\rangle
^2-\frac 12M_\pi ^2\left\langle \Pi \right\rangle ^2+\frac 12M^2\left\langle
\omega _\mu \right\rangle ^2+\frac 12M^2\left\langle A_\mu \right\rangle ^2
\label{pp}
\end{equation}
and the energy density 
\begin{equation}
\left\langle E\right\rangle =\bar E+T\left[ \left( \frac{\partial
\left\langle P\right\rangle }{\partial T}\right) _{V,\mu }-\left( \frac{%
\partial \bar P}{\partial T}\right) _{V,\mu }\right] +\frac
12M_s^2\left\langle s\right\rangle ^2+\frac 12M_\pi ^2\left\langle \Pi
\right\rangle ^2-\frac 12M^2\left\langle \omega _\mu \right\rangle ^2-\frac
12M^2\left\langle A_\mu \right\rangle ^2  \label{ee}
\end{equation}
where \cite{TV2011c,T2012b} 
\begin{equation}
\bar P=\frac \gamma {2\pi ^2}T\int\limits_0^\infty \ln \left( 1+\exp \frac{%
\mu -\left\langle \varepsilon _p\right\rangle }T\right) p^2dp  \label{p}
\end{equation}
is the pressure, while 
\begin{equation}
\bar E\equiv \mu n_5+T\left( \frac{\partial \bar P}{\partial T}\right)
_{V,\mu }-\bar P=\frac \gamma {2\pi ^2}\int\limits_0^\infty \left\langle
\varepsilon _p\right\rangle \,f_\varepsilon \,p^2dp  \label{e}
\end{equation}
is the energy density and 
\begin{equation}
n_5=\frac \gamma {2\pi ^2}\int\limits_0^\infty f_\varepsilon \,p^2dp
\label{n55}
\end{equation}
is the axial charge density of a free tachyon Fermi gas with distribution
function 
\begin{equation}
f_\varepsilon =\frac 1{\exp \left[ \left( \left\langle \varepsilon
_p\right\rangle -\mu \right) /T\right] +1}  \label{f}
\end{equation}
and single-particle energy spectrum $\left\langle \varepsilon
_p\right\rangle $ determined in the frames of mean-field approximation by
substituting the mean fields $\left\langle s\right\rangle $, $\left\langle
\Pi \right\rangle $, $\left\langle \omega _\mu \right\rangle $ and $%
\left\langle A_\mu \right\rangle $ in (\ref{ea}).

\section{Vector and pseudovector condensates}

Let us consider tachyons with pure vector and pseudovector interaction $%
\omega _\mu $ and $A _\mu $. For a rotational symmetric uniform matter we
take $\left\langle \vec \omega \right\rangle =\left\langle \vec
A\right\rangle =0$, $\left\langle \omega _0\right\rangle =\tilde \omega _0=%
\mathrm{const}$ and $\left\langle A_0\right\rangle =\tilde A_0=\mathrm{const}
$ so that equations (\ref{aa1}) and (\ref{oo1}) yield 
\begin{equation}
\tilde \omega _0=-\frac g{M^2}n  \label{o1}
\end{equation}
\begin{equation}
\tilde A_0=-\frac e{M^2}n_5  \label{a1}
\end{equation}
where, again, the coupling constants $g$ and $e$ are written explicitely,
and 
\begin{equation}
n_5=\left\langle \bar \psi \gamma ^0\gamma _5\psi \right\rangle  \label{o2}
\end{equation}
is the axial charge density, while 
\begin{equation}
n=\left\langle \bar \psi \gamma ^0\psi \right\rangle  \label{o3}
\end{equation}
is the particle number density (and the coupling constants $g$ and $e$ are
written explicitly).

Thus, the vector and pseudovector fields acquires finite condensed values $%
\tilde A_0$ and $\tilde \omega _0$ which depends on $n$. Substituting $%
\tilde \omega _0$ (\ref{o1}) in (\ref{ew}) we find the mean-field energy
spectrum of tachyons 
\begin{equation}
\left\langle \varepsilon _p\right\rangle =\sqrt{p^2+\frac{n^2g^4}{M^4}+hp%
\frac{ng^2}{M^2}-m^2}-eA_0  \label{o33}
\end{equation}
Defining effective single-particle energy 
\begin{equation}
\varepsilon _{*p}=\sqrt{p^2+\frac{n^2g^4}{M^4}+hp\frac{ng^2}{M^2}-m^2}
\label{o33a}
\end{equation}
together with effective chemical potential 
\begin{equation}
\mu _{*}=\mu +eA_0  \label{o33b}
\end{equation}
and taking into account that the distribution function (\ref{f}) can be
presented in the form 
\begin{equation}
f_\varepsilon \equiv \frac 1{\exp \left[ \left( \varepsilon _{*p}-\mu
_{*}\right) /T\right] +1}  \label{ffff}
\end{equation}
we calculate the pressure (\ref{pp}) and energy density (\ref{ee}) so

\begin{equation}
\left\langle P\right\rangle =\bar P+\frac 12\frac{g^2n^2}{M^2}+\frac 12\frac{%
e^2n_5^2}{M_A^2}  \label{pp9}
\end{equation}
\begin{equation}
\left\langle E\right\rangle =E_{*}+T\left[ \frac{e^2n_5}{M_A^2}\left( \frac{%
\partial n_5}{\partial T}\right) _{V,\mu }-\frac{g^2n}{M^2}\left( \frac{%
\partial n}{\partial T}\right) _{V,\mu }\right] -\frac 12\frac{g^2n^2}{M^2}%
+\frac 12\frac{e^2n_5^2}{M_A^2}  \label{ee9}
\end{equation}
where 
\begin{equation}
E_{*}=\frac \gamma {2\pi ^2}\int\limits_0^\infty \varepsilon
_{*p}\,f_\varepsilon \,p^2dp  \label{e9}
\end{equation}
is the energy density of free particles with energy spectrum $\varepsilon
_{*p}$ (\ref{o33a}) and chemical potential (\ref{o33b}).

The single-particle energy (\ref{o33a}) is given in Fig.\thinspace \ref{Fv1}
for right-handed and left-handed tachyons. The single-particle energy (\ref
{o33a}) of right-handed tachyons ($h=1$) is always real at any $p$ when 
\begin{equation}
n>n_{+}=\frac{mM^2}{g^2}  \label{o4}
\end{equation}
while the single-particle energy of left-handed tachyons ($h=-1$) is always
real at any $p$ when 
\begin{equation}
n>n_{-}=\frac 2{\sqrt{3}}\frac{mM^2}{g^2}  \label{o5}
\end{equation}
Thus, the pressure (\ref{pp}), (\ref{pp9}) and energy density (\ref{ee}), (%
\ref{ee9}) acquire no imaginary content if the particle number density
exceeds critical threshold $n_{+}$ (\ref{o4}) or $n_{-}$ (\ref{o5}).

The group velocity 
\begin{equation}
v=\frac{d\varepsilon _{*p}}{dp}=\frac{p\pm ng^2/\left( 2M^2\right) }{\sqrt{%
p^2+n^2g^4/M^4\pm png^2/M^2-m^2}}  \label{o6}
\end{equation}
associated with energy-spectrum (\ref{o33a}), is subluminal for both
right-handed and left-handed tachyons when inequality (\ref{o5}) is
satisfied. This group velocity is plotted in Fig.\thinspace \ref{Fv2}.

\textrm{What is important for us is be noted than in} the limit $%
m\rightarrow 0$ the energy spectrum () is 
\begin{equation}
\varepsilon _{*p}=\sqrt{p^2+\frac{n^2g^4}{M^4}\pm p\frac{ng^2}{M^2}}
\label{o33a}
\end{equation}
and the group velocity () is helicity dependendt 
\begin{equation}
v=\frac{p\pm ng^2/\left( 2M^2\right) }{\sqrt{p^2+n^2g^4/M^4\pm png^2/M^2}}
\label{o6}
\end{equation}
and It is given in Fig. 3 and Fig. 4.\textrm{\ Omega is incuded in the mass
matrix, so that the vector condnensate} results in the mass generation.

\section{Conclusion}

The energy dispersion relation (\ref{ea}) of a tachyon coupled to the
scalar, pseudoscalar, vector and pseudovector (electromagnetic) fields
reveal the nature of interaction. The scalar and pseudoscalar potentials
field contribute to the only effective mass (\ref{em}). The potential is
acting as a regular electromagnetic field on the ordinary Dirac particle.
The most interesting behavior belongs to the vector field, whose sole action
results in very peculiar energy spectrum (\ref{ew}) including dependence not
only on the spin but also on the helicity of the tachyon.

\begin{figure}[tbp]
\caption{Single-particle tachyon energy at small density (dotted), critical
density (\ref{o4}) or (\ref{o5}) (dashed) and large density (solid) vs
momentum (both in the unit of mass $m$) for right-handed (R) and left-handed
(L) tachyons.}
\label{Fv1}{\includegraphics[scale=0.75]{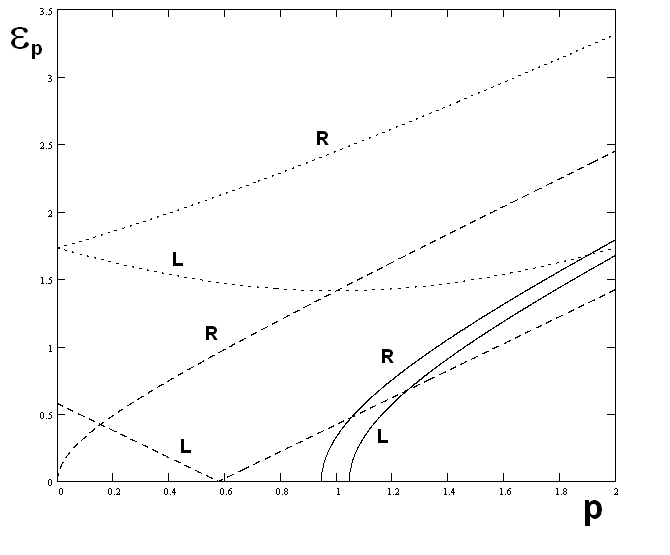}}
\end{figure}

\begin{figure}[tbp]
\caption{Group velocity of right-handed (R) and left-handed (L) tachyons at
small density (dotted), critical density (\ref{o4}) or (\ref{o5}) (dashed)
and large density (solid).}
\label{Fv2}{\includegraphics[scale=0.65]{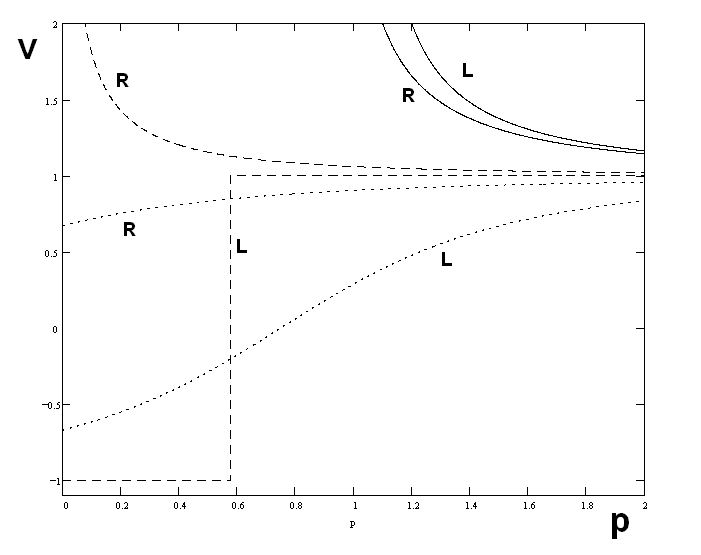}}
\end{figure}

Working in the frames of mean-field approximation, we find that the
single-particle energy (\ref{o33}), (Fig.~\ref{Fv1}) is real under
conditions (\ref{o4})-(\ref{o5}) that imply that the pressure (\ref{pp9})
and energy density (\ref{ee9}) are also real like that of an ordinary Fermi
gas. This means that interaction results in stabilization of the tachyon
matter, and weak coupling ($g\rightarrow 0$) can provide stability to the
dense matter, while strong coupling ($g\rightarrow \infty $) is enough for
stabilization of rarefied matter. It should be also noted that the group
velocity (\ref{o6}) becomes subluminal in dense tachyon Fermi gas, see
Fig.\thinspace \ref{Fv2}, although right-handed and left-handed tachyons
reveal very different behavior: for example, left handed tachyons may have
negative group velocity like the optical tachyons in nonlinear medium \cite
{O1}, and subluminal right-handed tachyons are faster than left-handed
tachyons at the same momentum.

The present analysis may give further ideas for solving applied problems.
For example, we have all necessary theoretical background (\ref{bi})-(\ref
{ea}), (\ref{z}) to consider superconductivity of interacting tachyon Fermi
system and find the relevant excitation spectrum. It should be emphasized
that conditions (\ref{o4})-(\ref{o5}) do not warrant automatically the
causality condition \cite{TV2011c}, which may impose further requirements to
the particle number density. According to (\ref{ee9}) the vector and
pseudovector (electromagnetic) interaction bring different
repulsive-attractive effects to the many-particle system, and formulas (\ref
{pp})-(\ref{f}), (\ref{o33})-(\ref{e9}) are can be applied to calculation of
the binding energy of tachyon matter and finding the point of gas-liquid
phase transition. Solving such problems, one should be careful with
expression of the Fermi momentum, bearing in mind the important fact that
formulas (\ref{o33})-(\ref{o6}) are written with the particle number density 
$n=\left\langle \bar \psi \gamma ^0\psi \right\rangle $, while the partition
function of a free tachyon Fermi gas must include the axial charge density $%
n_5=\left\langle \bar \psi \gamma ^0\gamma _5\psi \right\rangle $.

The is author grateful to Erwin Schmidt for inspiring discussions.

\newpage 

\newpage 

\end{document}